\documentclass[conference]{IEEEtran}

\pdfoutput=1
\usepackage{graphicx}
\usepackage{amsmath}

\ifCLASSINFOpdf
\else
\fi
\begin{document}
%
\title{Privacy Concerns Regarding Occupant Tracking in Smart Buildings}

\author{\IEEEauthorblockN{Ellis Kessler}
\IEEEauthorblockA{Virginia Tech \\ 
Department of Mechanical Engineering\\
ellisk1@vt.edu}
\and
\IEEEauthorblockN{Moeti Masiane}
\IEEEauthorblockA{Virginia Tech\\
Department of Computer Science\\
moeti@vt.du}
\and
\IEEEauthorblockN{Awad Abdelhalim}
\IEEEauthorblockA{Virginia Tech\\
Department of Civil Engineering\\
atarig@vt.edu}}


%


\maketitle

\begin{abstract}

Tracking of occupants within buildings has become a topic of interest in the past decade. Occupant tracking has been used in the public safety, energy conservation, and marketing fields. Various methods have been demonstrated which can track people outside of and inside buildings; including GPS, visual-based tracking using surveillance cameras, and vibration based tracking using sensors such as accelerometers. In this work, those main systems for tracking occupants are compared and contrasted for the levels of detail they give about where occupants are, as well as their respective privacy concerns and how identifiable the tracking information collected is to a specific person. We discuss a case study using vibrations sensors mounted in Virginia Tech's Goodwin Hall in Blacksburg, VA that was recently conducted, demonstrating that similar levels of accuracy in occupant localization can be achieved to current methods, and highlighting the amount of identifying information in the vibration signals dataset. Finally, a method of transforming the vibration data to preserve occupant privacy was proposed and tested on the dataset. The results indicate that our proposed method has successfully resulted in anonymizing the occupant's gender information which was previously identifiable from the vibration data, while minimally impacting the localization accuracy achieved without anonymization.

\end{abstract}

\vspace{5 pt}


%
\IEEEpeerreviewmaketitle

\section{Introduction}

\subsection{Motivation}

Tracking people in buildings has various implications including locating and accounting for people in case of emergency (e.g. lost children in a building) and  making sure all people are evacuated in case of a natural disaster. Other implications include identifying threats, making sure that unwanted people are prevented from being in restricted areas, maximizing occupant comfort, increasing energy efficiency \cite{molina-solana_data_2017} and collecting data for research e.g. collecting building simulation data \cite{yan_occupant_2015}.

Various mechanisms including the use of humans, cameras and electronic sensors have been used for tracking. Unfortunately, this tracking can lead to the gathering of information that can be used to violate the privacy and confidentiality of the people being tracked. The consequences of such breaches of information could lead to actual harm to individuals in cases where a malicious person can identify times when potential victims are vulnerable.

Confidentiality concerns occur when the individual being tracked does not want his or her whereabouts known, as could be the case if a celebrity is trying to get away from the public eye. Similar concerns also occur in cases of national security.  What is needed is a method for tracking while maintaining forward and backward privacy. Forward privacy and backward privacy prevent anyone with any data from being able to determine future tracking activity or being able to determine past tracking information from that data. Both of these concerns deal with occupant data anonymity.

\subsection{Problem statement}
Current popular occupant tracking systems gather large amounts of Personally Identifiable Information. In this study, we try to assess if vibration-based tracking can be used to achieve comparable tracking benefits without the added risk of security and privacy violation.

\subsection{The significance of the problem}
The current monitoring devices could have various negative effects on a building’s occupants. Such effects include, and are not limited to:
\begin{itemize}
\item Security vulnerabilities as a result on the growth in the deployment of Internet of Things (IoT) devices in building management and their inherent security concerns, as well as the collection and storage of data generated by these devices.
\item Privacy vulnerabilities that can result from personally identifiable information falling into the wrong hands.
\item Possible victimization of or bias against particular occupants who have a particular identifiable behavior.
\item Possible health impacts of wireless technology.
\end{itemize}

\subsection{Contributions}

In this work, we make the following contributions: 
\begin{itemize}
\item We present an evaluation of current ethical safeguards that are being used for occupant tracking.
\item We examine a novel tracking approach currently being used in Virginia Tech's Goodwin Hall, the most instrumented building in the world for vibrations.
\item We propose a method to preserve the anonymity of the tracked occupants while maintaining localization accuracy.
\end{itemize}

\newpage
\section{Related Work}

\subsection{The Development, Ethical and Privacy Concerns of Traditional Tracking Approaches}

Personal tracking technology has immensely evolved over the past three decades. With high demand from the government, security and commerce sectors, and massive technological advancements to make it all possible. Modern-Day personal tracking in the way we know it mainly became possible due to the advancements in the Global Positions System (GPS), originally known as Navstar GPS, a satellite-based radio-navigation system owned by the United States government and operated by the United States Air Force, which is a global satellite system providing the geographic location and time information to a GPS receiver anywhere on Earth. It was developed in the early 1970s by US Department of Defense to help track military troops and equipment, before becoming available for public use starting in the late 1980s for special use, 1995 with selective capabilities for public use and full capabilities since the year 2000.
The GPS does not require the user to transmit any data, and it operates independently of any telephonic or internet reception, though these technologies can enhance the usefulness of the GPS positioning information, making it a cheap technology to acquire even at its infancy. GPS trackers came to public higher use for personal trackers in the late 90s, early 2000s, before becoming actively used in modern day mapping and navigation systems.
As GPS use began to become mainstream, privacy concerns regarding the use of GPS tracking begun to arise. An excellent early study was conducted by Waseem Karim \cite{karim1}, who shed the light on the emerging privacy concerns and possible violations of the 4th Amendment of the US Constitution (protection against unreasonable search) through the data gathered by the available GPS trackers at the time that would be readily accessible to Law Enforcement, especially with new post 9/11 legislations at the time such as the Patriot Act and the Federal Communications Council requiring all cell phone providers to equip new cellphones with GPS receivers. Similar studies were later conducted by K Michael et. al \cite{michael2} and J Wang et. al \cite{wang3}, who looked into the different emerging applications of personal tracking systems and their uses in tracking suspects, employees and children, and the ethical implications associated with that.

\begin{figure}[!h]
\centering
\includegraphics[width = 3.4in, height = 2in]{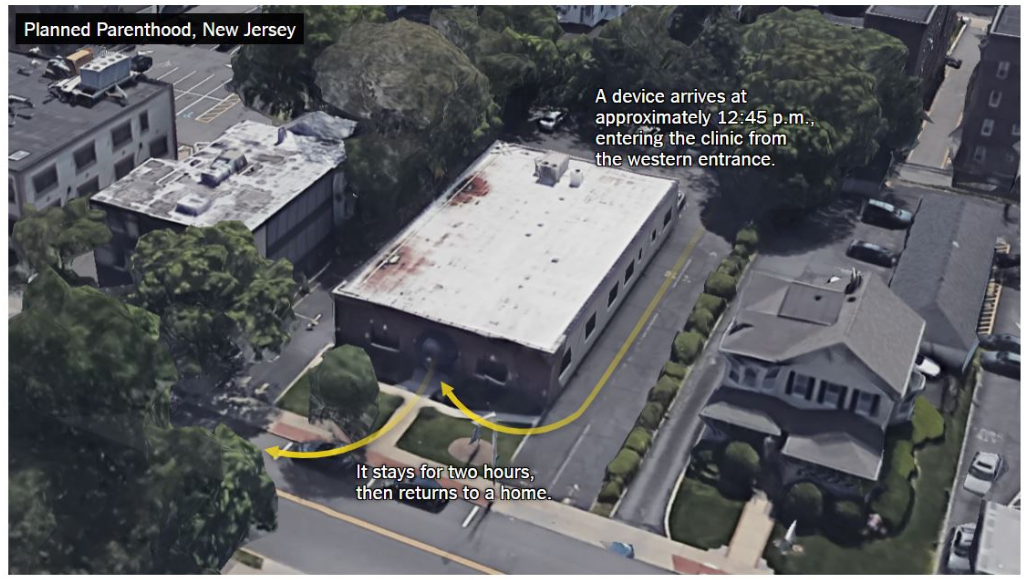}
\caption{\label{fig:gpstrack} Sample of the high-quality tracking data obtained and investigated by the New York Times.}
\end{figure}

The early red-flags raised by the above researchers and many more, however, have been taken for granted. A recent study conducted at the Northeastern University in Boston \cite{cnbc5} has concluded that mobile phones, with all the sensors, are in essence the best spying device one could imagine. The researchers developed a simple flash-light application that when installed, acquires access to the user’s phone’s GPS sensors. The researchers also demonstrated how the user restrictions on location data can be easily over-passed by software developers. Another recent report by the New York Times \cite{nyt4} highlights the way applications access, collect and commercialize personal location data from user’s phones at an industrial scale. The reporters accessed data from an undisclosed source (hinted as a major mobile phone application provider) has shown how an individual’s location was recorded with high-quality location and timestamps over 8,600 within 4 months, almost once every 21 minutes. The data was of granularity high enough to identify the individual’s place, time spent, travel speed, and common and uncommon visits as clearly demonstrated in Figure \ref{fig:gpstrack}. The accuracy of the location was found to be within a few yards, and for some users, data was collected over 14,000 times a day. And while the location data is device data and not personal data, the high granularity allows identifying individuals home and work locations, which in combination with personal data that may be acquired from other sources leads to an unprecedented, on-the-hour privacy violation of individuals.

\begin{figure*}[!t]
\centering
\includegraphics[width = 7in, height = 3 in]{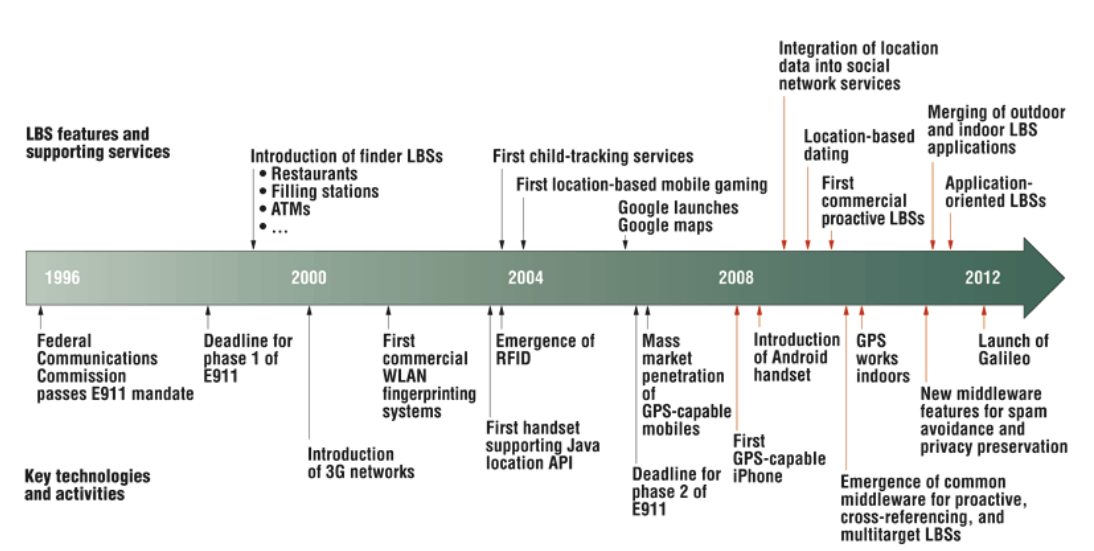}
\caption{\label{fig:timeline} Timeline for the development of Location-Based tracking systems.}
\end{figure*}

Surveillance cameras have been around since the late 80s. The main applications were around-the-clock monitoring of sensitive areas and buildings, and traffic violations. Later in the 90s cameras were introduced to monitor ATMs. It was only until 1996 that the first IP linked camera was released, meaning that video recording from surveillance cameras is no longer centralized, and recordings from multiple cameras are easily collected and aggregated in one place. An early study by Cristopher Slobogin \cite{slobogin6} highlighted the privacy concerns associated with the right to anonymity with the then-increasing use of surveillance cameras in public spaces. The growing use of surveillance cameras and massive advancements in computing and computer vision allowed for aggregating the centralized data from multiple sources and tracking individuals with very high accuracy. As the privacy concerns with the vision-based tracking began to arise, researchers have placed great emphasis on preserving the benefits of high-quality surveillance while limiting the amount of privacy intrusion associated with that. An example of such effort was that by M Saini et. al \cite{saini7} who proposed a model to quantify privacy loss through surveillance footage. Adam Shwartz \cite{schwartz8} concluded that the city of Chicago’s highly integrated network of 20,000 surveillance cameras violates privacy, encourages racial profiling, and does not necessarily make the city any safer.

\begin{figure}[!h]
\centering
\includegraphics[width = 3.4in, height= 2.4in]{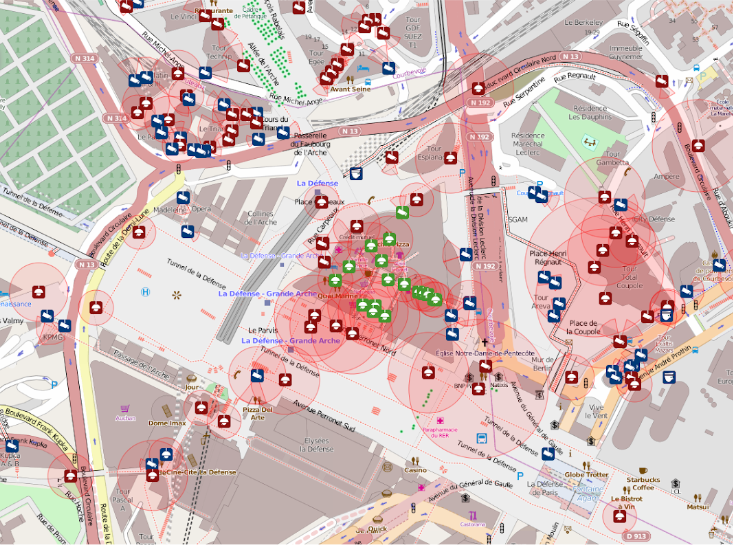}
\caption{\label{fig:cammap} A crowd-sourced open map showing the distribution of surveillance cameras around the Arc de Triomphe in central Paris.}
\end{figure}

The rapid development of artificial intelligence object recognition and tracking techniques coupled with the already-mature location-based tracking technology has taken vision-based tracking into a completely new level. High-quality location data alongside images and videos from surveillance cameras, social media application and smart phones provide a wealth of information that may be used to improve public safety but at the same time provides a constant threat to the individuals’ privacy. Now a days in highly crowded areas, surveillance systems are used at an industrial scale. An example from central Paris is shown in figure \ref{fig:cammap}. In recent studies \cite{chan9}, \cite{wang10}, researchers proposed applications of existing artificial intelligence methodologies to offer innovative means of anonymizing vision-based tracking data to maintain the sought-after benefits while preserving the privacy of individuals within the monitored area.

The traditional location-based and vision-based tracking methods succinctly discussed in this section, however, under-perform when used indoors. The obstructions caused by buildings walls and floors significantly hinders the range of surveillance and the quality and accuracy of GPS signals. The focus of this study is to assess how those short-comings of traditional tracking approaches in terms of privacy concerns and under-performance indoors can be addressed by using vibration-based sensors augmented with state-of-the-art data analytics approaches.

\subsection{Current Indoor Tracking Methods}

\def\arraystretch{1.125}
\begin{table*}[!t]
    \caption{\label{tab:VibrationBased} Summary of various vibration based localization methods in the literature. Methods are presented with the information they require, and the error reported in the work.}
    \begin{center}
\begin{tabular}{r l c c c c c}

Localization Based On & Method & \# Sensors & Type & Information Needed & Error (m) & Ref \\
\hline
Energy & Weighted Sensor Coordinates & 6 & A & Event Energy & 0.85 & \cite{alajlouni_new_2019}\\ 
  & Energy Attenuation over Distance & 4 & A & Event Energy & 0.25 & \cite{alajlouni_impact_2018} \\
  \\
  Time & Time Difference of Arrival (TDoA) & 12 & A & Time of Arrival & 0.70 & \cite{poston_indoor_2017,poston_towards_2015} \\
  & Sign Only TDoA & 9 & A & Time of Arrival & 0.46 & \cite{bahroun_new_2014}\\
  & Wavelet Transformed TDoA & 9 & G & Time History & 0.34 & \cite{mirshekari_occupant_2018}\\
  \\
  Dispersion Compensation & Warped Frequency Transform Correlation & 15 & A & Time History & 0.15 & \cite{woolard_applications_nodate}\\
\end{tabular}
\end{center}
\end{table*}

Current indoor tracking tracking techniques \cite{mainetti2014survey} \cite{basri2016survey} fall into two categories: i) wireless and ii) vision. Wireless technologies include GPS \cite{van2001indoor} \cite{soloviev2011extending}, Ultrasound \cite{knauth2012iloc+} \cite{cote2013indoor}, RFID \cite{guerrieri2006rfid} \cite{bouet2008rfid}, WLAN \cite{shin2010wi} \cite{shrestha2013deconvolution} and Bluetooth \cite{fischer2004bluetooth} \cite{zhuang2016smartphone}, while vision technologies include fixed and mobile camera based tracking. Wireless technologies are characterized  by a transmitter that sends a radio signal and a receiver on the other end. The general intuition is that the receiver makes a determination of the location of the transmitter based on how much time it takes for the signal to travel between the transmitter and receiver, or by measuring the signal strength at the receiver. Due to various challenges including attenuation and packet loss that affect radio signals \cite{adib2013see}, various techniques like addition of redundant receivers and transmitters as well as the manipulation of the signal strength are employed. Additional considerations for wireless tracking include power management for the receiver and transmitters. Vision based technologies also have similar vulnerabilities like the need for line of sight for fixed cameras as well power management considerations for mobile cameras. Even though wireless and vision based tracking techniques have been shown to adequately track building occupants, they present a concern in terms of maintaining privacy for building occupants. Data supplied by a camera, or cell phone or RFID badge assigned to a building occupant for example, can be used to determine the location and activities of a subject being tracked and as a result violate their privacy. This work on the other hand, utilizes building sensors that are not associated with particular building occupants and as a result do not suffer from these vulnerabilities, a topic that \cite{cascone_ethical_2017,sharpe_ethical_2019} have touched on.     

\begin{figure}[!h]
\centering
\includegraphics[width = 3.4in, height =1.6in ]{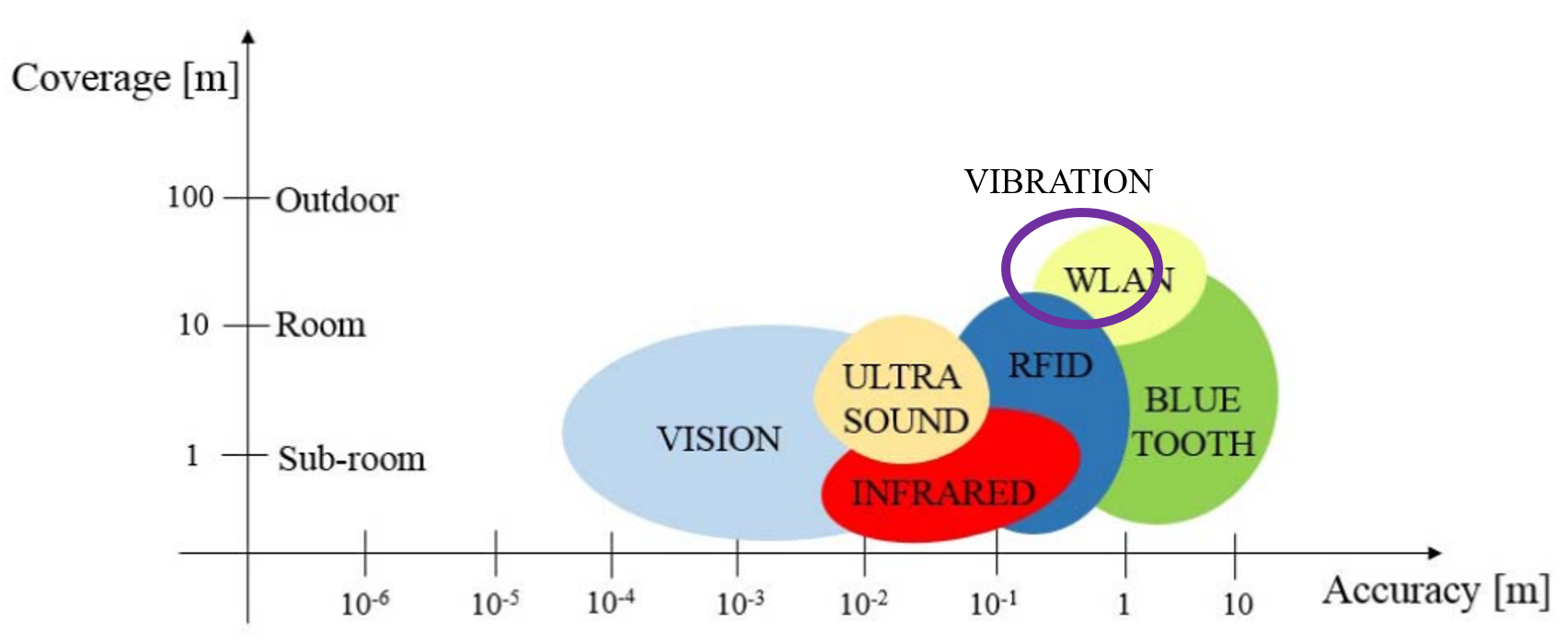}
\caption{\label{fig:Localization_Comparison} Coverage and accuracy of various indoor localization methods, adapted from \cite{mainetti2014survey}. An ellipse line has been added to represent where vibration based tracking would compare with these other methods.}
\end{figure}

\subsection{Vibration-Based Localization in Buildings}


Multiple methods for localizing events--such as a footstep--in a building using vibration sensors have been demonstrated. These methods can be broken into three main groups: energy based methods, time based methods, and dispersion compensation based methods. Each of these groups may have various modifications, while still relying on the same underlying principle. Energy based methods rely on the attenuation of the signal from an event as the vibrations travel a distance \cite{alajlouni_impact_2018}. Comparing the signal energy at multiple sensors allows the original location of the event to be predicted. Time based methods records the time that vibrations reach an array of sensors, and through knowledge of the wave propagation speed in the material of the floor, the location of the event is predicted through multilateration. Since the exact time of the event is often unknown, differences in time of arrivals are often used. For this reason time based methods are usually referred to as Time Difference of Arrival (TDoA) \cite{poston_indoor_2017}. While TDoA methods are reliable with waves propagating through media such as air, vibrations in a solid structure present another hurdle. Wave propagation in solids exhibits a phenomenon called dispersion, meaning different frequency waves travel at different speeds. To overcome this hurdle, modifications to TDoA such as using wavelet transforms to track only a single frequency have been demonstrated \cite{mirshekari_occupant_2018}. Another method for localization is to leverage the dispersion phenomenon itself. In a group of methods we will refer to as dispersion compensation, the dispersion relation of the floor is used to transform the time domain signal to a ``distance" domain. Methods for this transformation such as the Warped Frequency Transform (WFT) then allow differences in distances from an array of sensors to predict the location of the event \cite{woolard_applications_nodate}. Although only one dispersion compensation method has been demonstrated in a building, these methods are common in smaller structures such as plates to detect damage \cite{perelli_passive_2012,wilcox_rapid_2003,cai_distance-domain_2017}. Table \ref{tab:VibrationBased} shows a summary of these localization methods along with the number sensors used in each study, the sensor type used, the information needed, and the error reported.

The different amounts of information needed for each localization method shown in Table \ref{tab:VibrationBased} come with different levels of privacy concerns. Methods such as WFT and wavelet transformed TDoA require a full time history of the event in order to localize it. Other time based methods also require the time of arrival from multiple sensors, which in practice would require the full time history of the event. With the shape of the vibration time history of an event such as a footstep, it has been shown that the gender of the person can be determined as well as distinguishing between individuals \cite{bales_gender_2016,pan_footprintid:_2017}. Therefore, both dispersion compensation and time based methods present a moderate privacy risk. Although some distinguishing information about an individual could be extracted from these methods, this information is still far less distinguishing than an image or a wireless device signal. Still, the energy based tracking methods show promise to provide even less privacy risk. The only feature needed for this type of localization is the signal energy during an event. In practice, this has been accomplished using the entire time history by detecting the event and calculating the signal energy over a certain period of time. Using this method, no more privacy is ensured than the other vibration based methods. However, we propose an amendment to the procedure to calculating energy of events which will allow more privacy retention. Instead of storing a full time history, the time series was broken up into windows over which the signal energy is calculated. Storing only the energy over windows is hypothesized to anonymize the data by obscuring the features in the time history which could be used to identify an individual, while keeping the data needed to localize the event.

\section{Approach and Experiments}

Two main experiments have been used to examine the accuracy and privacy concerns regarding using vibration sensors for tracking occupants in a smart building. Data from an evacuation due to a fire alarm has been used in the first experiment to demonstrate the ability of vibration sensors to track both group movement, as well as individual movement within a building. The precision of this tracking method has also been compared to conventional indoor tracking methods. A second experiment has been used to investigate the level of personal identification possible with only vibration sensors. This experiment was carried out on a dataset containing both male and female subjects walking down a hallway. First, the ability to classify subjects into groups based on sex, or to identify individuals was examined with raw data. Second, the same classifications were attempted after obscuring the data by reducing it to only measure the amount of energy over windows of time. This energy windowed data was also be used in the vibration based tracking method, showing the benefits of indoor tracking with less personal identifying information measured.

\subsection{Goodwin Hall and Data Pre-Processing}

\begin{figure}[!h]
\centering
\includegraphics[width = 4in]{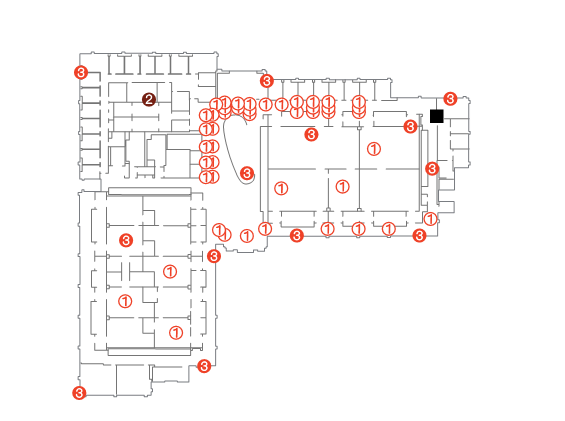}
\caption{\label{fig:GoodwinLayout} An example of the sensor layout on the fourth floor of Goodwin Hall. The numbers indicate how many directions have a sensor measuring at that location. The locations with one sensor measure in a single direction (vertical), while locations with three sensors measure in the vertical and both horizontal directions.}
\end{figure}

Goodwin Hall on Virginia Tech's campus is the most instrumented building in the world for vibration. Goodwin Hall has 225 high sensitivity accelerometers permanently mounted to the building's structure. Figure \ref{fig:GoodwinLayout} shows an example of a sensor layout from the fourth floor. These sensors are connected to one of five data acquisition boxes, which are in turn connected through one of two switches to a central server where data is stored. There is an established data pipeline within the Virginia Tech Smart Infrastructure Laboratory (VTSIL) which converts raw data from the server in Goodwin Hall to time series data in either .mat (Matlab data file) or .csv files for ease of use. The coordinates of each sensor in an \textit{x-y-z} coordinate system are also embedded in this pipeline, so that sensor locations are automatically available in the same order as time histories from each sensor. The system integrated in Goodwin Hall is capable of recording data at a variety of sampling frequencies up to around 50 kHz.

\begin{figure*}[!t]
\centering
\includegraphics[width = 7in]{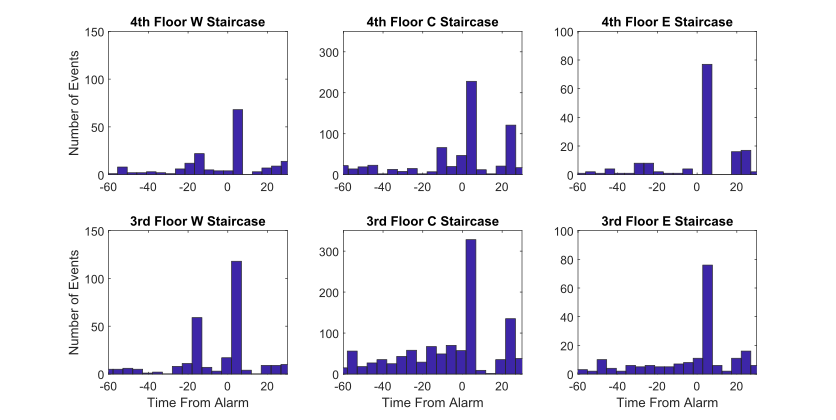}
\caption{\label{fig:ExitCounter} Number of events recorded at sensors near stairwells on floors 3 and 4 before, during, and after evacuation. The number of events in 5 minute intervals are added and reported. The time is measured in minutes from the beginning of the fire alarm.}
\end{figure*}

As mentioned, there is a data pipeline to import raw data from the building taken at any sampling rate into a readable format. After importing the raw data, there were three main pre-processing steps to be taken. First, the data had to be de-trended. The accelerometers sometimes take time to reach their charged state, as well as the building being a common place for low frequency noise. To eliminate these drifts, the data was de-trended using a moving average filter with a window of one second. This method of de-trending has become commonplace in the VTSIL methodology. Next, some sensors were disregarded. There are a few places in the building where sensors are mounted close to a high noise source, for example near the HVAC ventilation fans. These sensors do not contain useful information, and are therefore removed before further analysis. Finally, the data is clipped so that only times of interest are kept. For the fire alarm dataset, this meant separating time before the evacuation, during the evacuation, and after the evacuation. This time clipping is easily performed because the data pipeline encodes the data file with the time that the recording began. Using this time, the sampling frequency, and the time when key events (such as the fire alarm sounded) the data can be clipped. Another type of time clipping is necessary before classification of footstep responses can be performed. In order to compare two footsteps to decide if they came from the same person, or are the same sex, the footsteps must be lined up in time somehow. In this work, the response was kept for a certain amount of time before the footstep was detected and a longer time after the footstep was detected. For example, 0.1 seconds may be kept before the detection, and 0.3 seconds after the detection. This ensures that footsteps responses are always vectors of the same length so that machine learning algorithms may be applied to classify the footsteps.

\subsection{Fire Alarm Dataset}

One dataset which was used in this work is that of all building occupants evacuating the building during an unplanned fire alarm. During this time, the building was already set to record ambient vibrations at a sampling rate of 256 Hz. This dataset has been used to show the possibility of using vibration sensors to track footsteps and groups of people as they move through a building. Over the course of about 10 minutes, all occupants of the building have been shown to travel toward exits the exits by increased footstep events detected by sensors closer to the exits. For about 20 minutes while the building is evacuated, there were no footstep events in the building. This dataset was used to show the benefits of indoor tracking. Individuals were tracked while evacuating, showing their evacuation route. This could lead to finding underutilized evacuation routes, or choke points which would slow down a future evacuation.

\begin{figure} [!h]
    \centering
    \includegraphics[width = 3.5in]{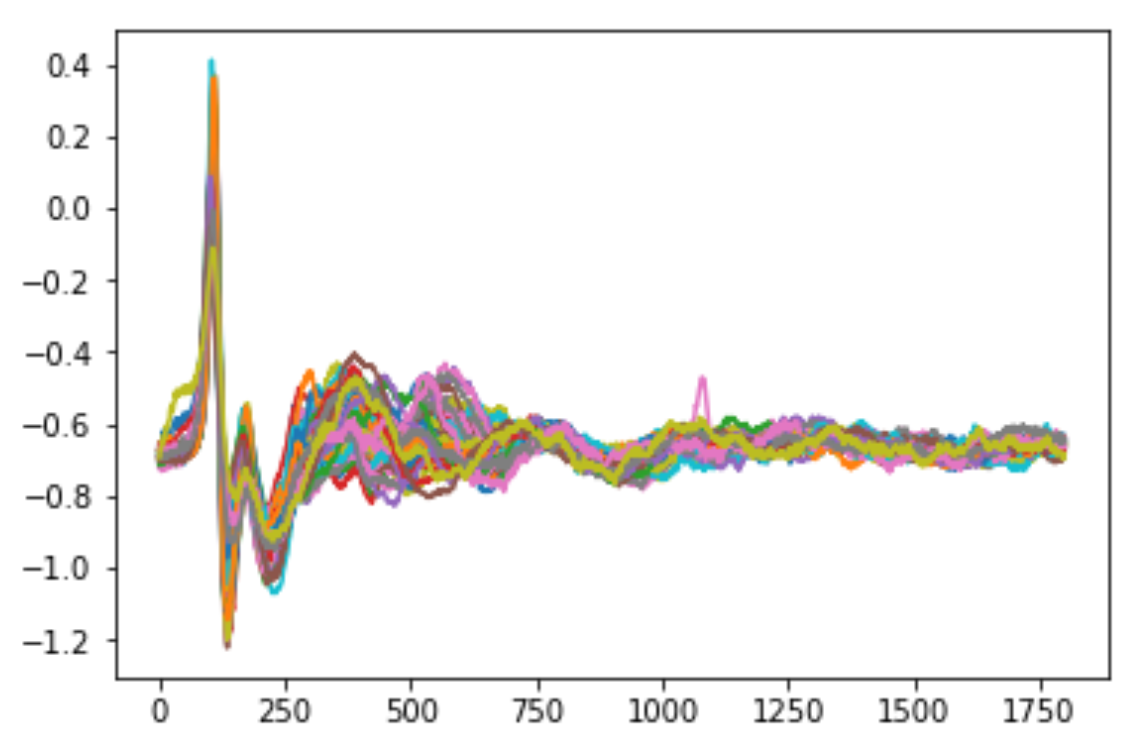}
    \caption{Single instance for acceleration readings during the fire alarm situation for multiple individuals.}
    \label{fig:FireAlarmTimeSeries}
\end{figure}

\subsection{Individual People Walking Dataset}

A second dataset was used which has data of single participants walking down a hallway on the fourth floor of Goodwin Hall. It is labelled with the sex of each participant, along with which trial belongs to which participant. This dataset was utilized to explore how personally identifying vibration data is. Previous works have shown that sex can be classified based on vibration data \cite{bales_gender_2016}, as well as preliminary indications that individuals can be identified by their vibration signal induced by walking \cite{pan_footprintid:_2017}. In this work, the amount of personal identifying information was compared before and after obscuring the data with a energy windowing process. It is hypothesized that both classification of sex as well as individual identification will be less accurate after obscuring the data. Multiple classification methods were used to classify either sex or try to identify individuals. These classification methods were evaluated based on their accuracy, and reductions in accuracy after obscuring the personally identifying information verify that less privacy violations are possible with said data.

\section{Data Analysis and Results}

\begin{figure*}[!h]
\centering
\includegraphics[width = 7in, height= 3in]{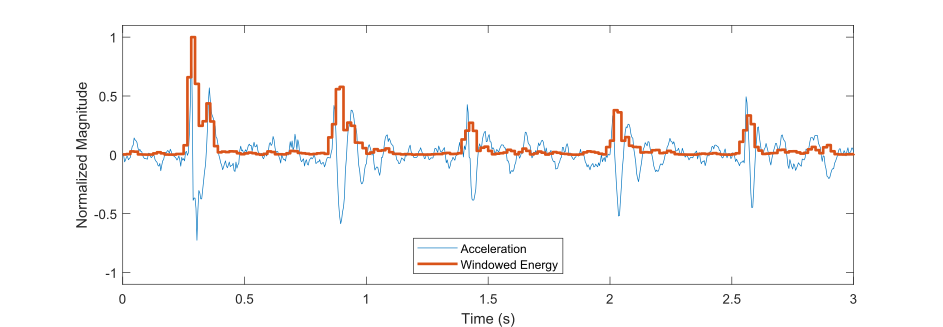}
\caption{\label{fig:EnergyWindow} Example of an acceleration time history, along with the corresponding windowed energy.}
\end{figure*}

\subsection{Data Exploration}

First, an exploratory data analysis of the fire-alarm dataset has been performed. Sensors were selected near the stairwells on the third and fourth floors. Methods similar to the energy windowing discussed previously were used to detect events with high signal energy. All events over a two hour period containing the evacuation were detected. Figure \ref{fig:ExitCounter} shows the number of events happening near each sensor over the course of five minute intervals before, during, and after the evacuation. Soon after the alarm rings, there are significant spikes in the number of events near each sensor, indicating that the building's occupants are evacuating. After the evacuation, there is a time of very few events happening since there should be no people left in the building. After about 20 minutes, events begin to increase again as people are let back into the building. 

\begin{figure}[!h]
\centering
\includegraphics[width = 3.5in, height = 2in]{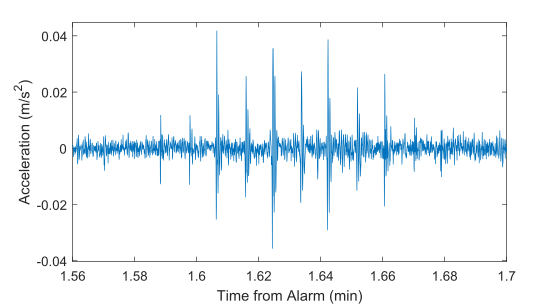}
\caption{\label{fig:SingleEvacuation} Walking signal generated by a single person, as measured by one sensor.}
\end{figure}

Figure \ref{fig:ExitCounter} demonstrates the capability to track activity near a certain sensor. It is also a goal of this project to track the location of occupants during the evacuation, which could be used to check paths of evacuation or build heatmaps of areas which are more trafficked. To this end, a signal which appears to be a single person walking through the hallway has been located in the fire alarm dataset. Figure \ref{fig:SingleEvacuation} shows the walking signal of a single person walking closer to a sensor, and then walking away from it. The amplitude of the signal can be seen to increase as the person gets closer, then decrease as the person gets further away. Using the energy calculated from a single footstep in this signal, as well as the energy from other sensors in the area, each footstep event can be localized. By localizing a sequence of footstep events, a single person could be tracked and the route they use for evacuation could be determined. Tracking individuals using this method, similarly to the energy based methods in Table \ref{tab:VibrationBased}, was demonstrated as part of the final report for this project.

The next step was to look at the feasibility of gender classification, and how to anonymize the gender if it is classifiable. The proposed method for anonymization is to break the data into windows, and calculate an \textit{rms} over each window. This had the effect of both downsampling the data, and significantly obscuring the shape of the response which could theoretically be used as personally identifiable information. An example of this energy windowing is shown in Figure \ref{fig:EnergyWindow}, as well as an equation for the windowing in Equation \ref{SNR}. The energy \textit{E} over a window of length \textit{w} is calculated by taking the square root of the sum of the squares of the acceleration data \textit{x}:
\begin{equation}
    E[k] = \sqrt{\frac{1}{w}\sum_{k=0}^{w-1}x^2[k]}
    \label{SNR}
\end{equation}

\subsection{Classification and Gender Anonymization}

An initial exploration of the data at each censor showed a damped harmonic oscillation behavior as shown in Figure \ref{fig:FireAlarmTimeSeries}. There is a huge sensor reading that coincides with the initial impact of the foot to the floor. This initial reading is followed by subsequent smaller readings due to the rest of the foot rolling over the ground. The data has positive and negative values showing the measured acceleration of the floor. The sensors measure the elastic effect of the ground as it reacts to the pressure of the foot. This results in the positive and negative sensor readings.

The main goal of this work is to preserve participant privacy while maintaining localization. As a proof of concept of our main goal we obscure the sex of our participants, while maintaining the ability to pinpoint their location in a building. We approach obscuring the sex of participants as a classification problem. Given an accurate model that can predict the sex of an individual walking in a building, using their footstep signature, can one reduce the accuracy of the model to that of a random classifier, while maintaining the ability to localize. To do this, we train several binary classification models and evaluate them using leave one out cross validation. We proceed to using increasing window sizes to aggregate the data in an effort to find a threshold where the bin size reduces the accuracy of the classifier to 50\%. We then determine if we can provide localization using that bin size.

\begin{figure}[!h]
    \centering
    \includegraphics[width = 3.5in, height= 2.5in]{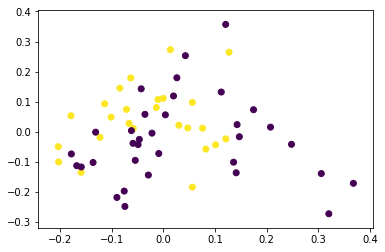}
    \caption{Principal components scatter plot.}
    \label{fig:pca}
\end{figure}

Our initial approach was to try a linear classifier. Using Principal Component Analysis (PCA), we reduce our data from 1800 dimensions, where each dimension is a reading from a sensor that has been sampled at a uniform rate of 4096 Hertz, to two dimensions. In Figure \ref{fig:pca} plot our data using a scatter plot. The scatter plot of principal components clearly indicates that the data is not linearly separable. This shows that we need more advanced classification approaches. We apply five widely used algorithms and train five supervised classifiers, namely a Neural Network, K Nearest Neighbours, Support Vector Machine, Decision Tree and Naive Bayes using the Scikit-learn Python Library. We use leave one out cross validation to tune hyper parameters for each model.

\begin{figure}[!h]
    \centering
    \includegraphics[width = 3.5in, height= 2.5in]{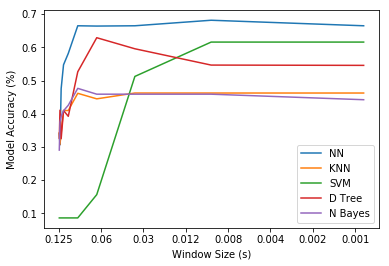}
    \caption{Classification accuracy with varying window size.}
    \label{fig:class_actual_full}
\end{figure}

The aforementioned models were developed for the dataset and our proposed energy-widowing approach was utilized. The effect of aggregating our sensor data into increasing window sizes is shown in Figure \ref{fig:class_actual_full}. The results show that aggregating and averaging the energy responses over bigger window size (lower sampling rate) significantly helps in reducing the classification accuracy for individual's gender.  For the Neural Network model, we were able to obscure information that has resulted in reducing its classification accuracy from about 70\% to below 50\% by using a window size of about 0.125 seconds. These results confirm our underlying hypothesis that aggregating the sensor data into windows of increasing size, would degrade the performance of the binary classifier to below 50\%.

\begin{figure}[!h]
    \centering
    \includegraphics[width = 3.5in, height= 2.5in]{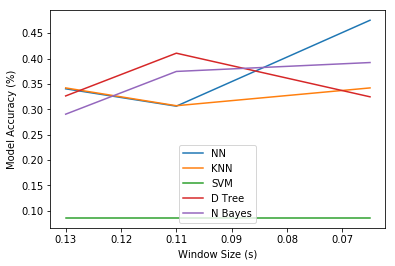}
    \caption{Classification accuracy with varying window size.}
    \label{fig:class_actual_part}
\end{figure}

However, for some models we see some behavior for large window sizes that seems to show an abrupt increase in accuracy, and this does not align with our initial intuition. It is hypothesized and expected that the accuracy would fall continuously with an increase in window size. To explore this further, we zoom into the area of the large window size in Figure \ref{fig:class_actual_part}. We believe this counter-intuitive behaviour is due to the high variance that would be present in minuscule dataset that we have used, where we only have responses for 16 participants. To assess this hypothesis, we used the data for the 16 participants to create a synthetic dataset.

We utilized Singular Value Decomposition (SVD) to identify the attributes that explain the variance in sensor acceleration within different participant gender. A process of trial-and-error demonstrated that using linear combinations of the first-two singular values alongside 23 other scaled singular values successfully generates an accurate representation of the actual dataset. This linear combination was used to synthesize data for 6,000 participants, of which 3,000 were females and 3,000 are males.

\begin{figure}[!h]
    \centering
    \includegraphics[width = 3.5in, height= 2.5in]{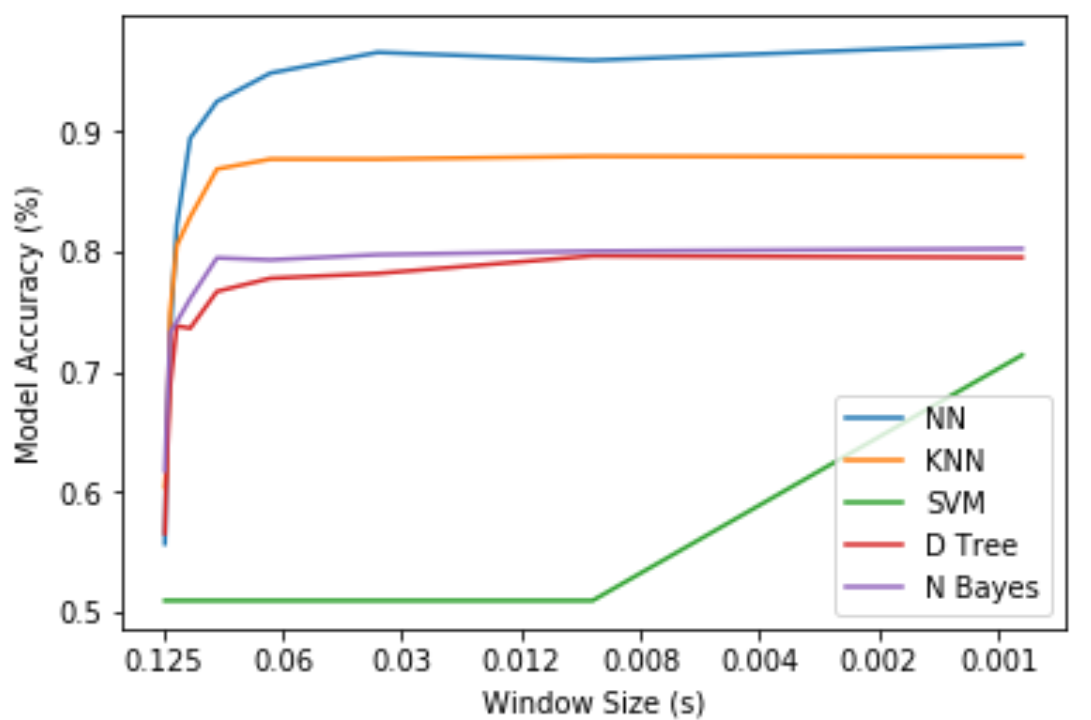}
    \caption{Classification accuracy with varying window size.}
    \label{fig:class_synthetic_full}
\end{figure}

\begin{figure}[!h]
    \centering
    \includegraphics[width = 3.5in, height= 2.5in]{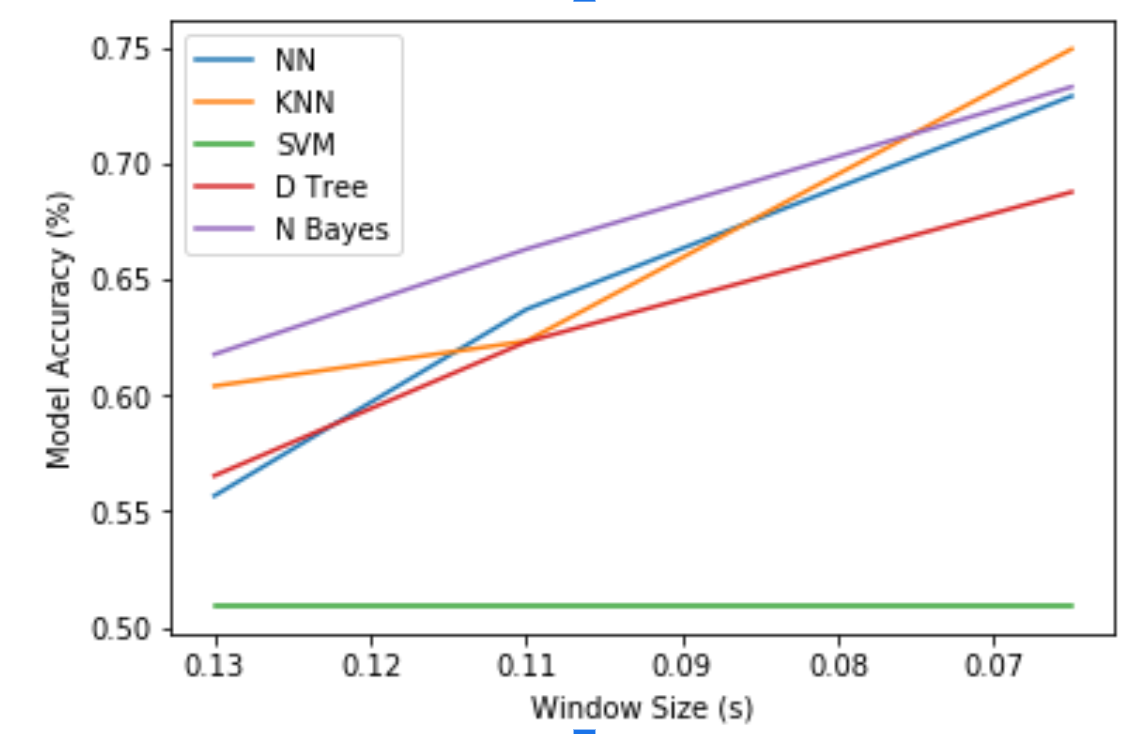}
    \caption{Classification accuracy with varying window size.}
    \label{fig:class_synthetic_part}
\end{figure}

Using our synthetic data, we see that the accuracy smoothly decreases with increased window size. This confirms that the abrupt increase in accuracy with larger window size was indeed an artifact of the small dataset that we have used. The very steep decrease in accuracy shown in figure \ref{fig:class_synthetic_full} is better explained when looking at the zoomed figure \ref{fig:class_synthetic_part}. We conclude that the increase in accuracy with the increase in window size can be ignored. Using our synthetic dataset, we also see a decrease in classifier accuracy with an increase in window size, even though this was not the point of generating the synthetic dataset.

\subsection{Occupant Localization}

In the previous section it has been shown that the energy windowing technique is successful at anonymizing identifying information in the floor's vibration signal. The next step, is to show that with the same window sizes, localization is still possible. In order to localize an event, the event must be detected in the signal first. This is done by finding peaks in the vibration energy signal. For example, there are five events in the signal in Figure \ref{fig:EnergyWindow}. Once an event is found in the time signal, the energy of that event at all sensors can be found by adding up the windowed energy from the time the event begins to a certain amount of time afterwards. This process is very similar to the process outlined by Alajlouni et. Al. \cite{alajlouni_impact_2018}, with the only modification that the original signal has been windowed before finding the energy of the event. The energy, $E_i$, at each sensor $i$ (located at position $x_i$ and $y_i$) from sensor 1 to sensor $N$ can then be used to find an estimate of the location of the event. This is done by using a non-linear least squares fitting of the energies at each location to the equation:

\begin{equation}
    E_i = E_s e^{-\beta |r_i|}
\end{equation}

where there are four unknowns: the original source energy $E_s$, the exponential attenuation factor $\beta$ for how the energy of the event decreases with distance, and the $x$ and $y$ coordinates of the event ($x_s$ and $y_s$) respectively, which are contained in the formula for the distance from the source to each sensor:

\begin{equation}
    r_i = \sqrt{(x_s - x_i)^2 + (y_s - y_i)^2}.
\end{equation}

\begin{figure}[!h]
\centering
\includegraphics[width = 3.5in, height= 2.5in]{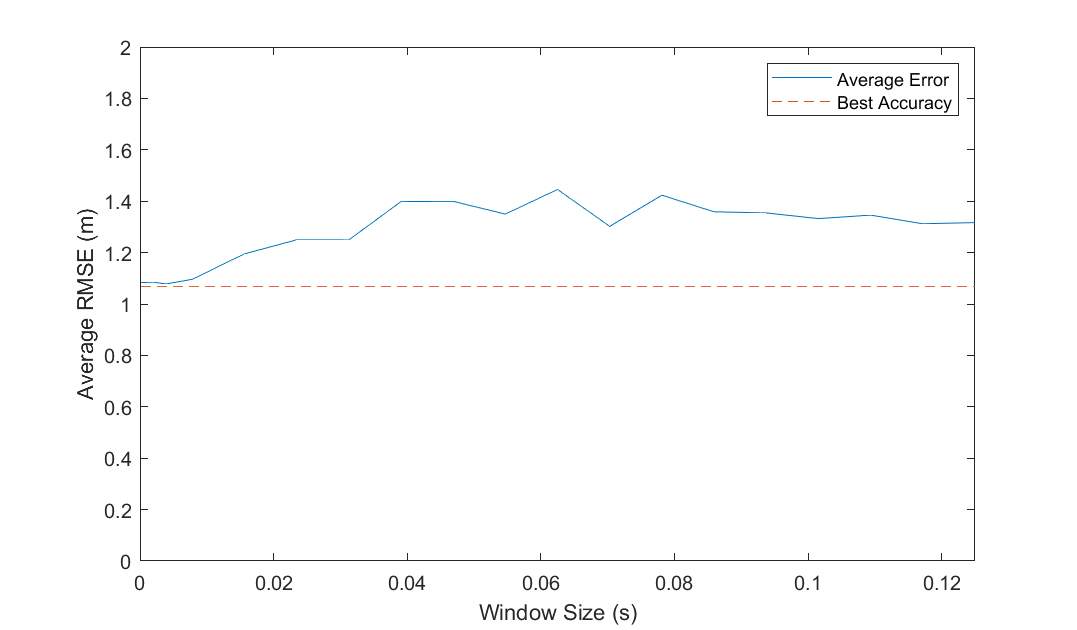}
\caption{\label{fig:LocalizationAccuracy} Average localization accuracy with various energy window sizes. The best accuracy for any single trial is also plotted as a constant for reference.}
\end{figure}

The underlying assumption of this localization method is that the energy decays exponentially with the distance away from the source, which has been shown by Alajlouni et. Al. \cite{alajlouni_impact_2018,alajlouni_new_2019}. Localization using this method was performed on a subset of the same dataset from the classification experiment. Once again, multiple different window sizes were used, this time examining the root mean squared error (RMSE) of the predicted event locations with known locations as the individual walks along a hallway. Figure \ref{fig:LocalizationAccuracy} shows how the RMSE changes as the window size for the energy windowing increases. The best localization accuracy in this experiment is around 1.1 m, which is not far off from the literature for vibration based localization techniques shown in Table \ref{tab:VibrationBased}. It is worth noting that the error in this work is slightly higher than that in the literature though, for which there are two possible explanations. First, energy based methods are not the most accurate to begin with, especially without some optimization of the parameters responsible for attenuation \cite{alajlouni_impact_2018}. A second reason for error is that the localization in this work was done over a large area. Events were localized all the way down a hallway for a length of over 100 feet, which is much larger than any single dimension spanned by events in any works in Table \ref{tab:VibrationBased}. Over these distances, the floor will begin to respond differently and it adds another level of complexity for a single method to work across the whole distance. Despite these limitations, Figure \ref{fig:LocalizationAccuracy} gives a good idea of how the localization error changes with changing window size. As the window size increases up to a maximum of 0.125 seconds (the largest size used in the classification experiment), the localization error only goes up from about 1.1 m to 1.4 m. A final localization accuracy of 1.4 m is still acceptable, it is more than accurate enough to localize someone to a specific room or a specific part of the building. Further, with multiple events in a row it was possible to get an even better idea of where a person is inside the building.

\begin{figure}[!h]
\centering
\includegraphics[width = 3.5in, height= 2in]{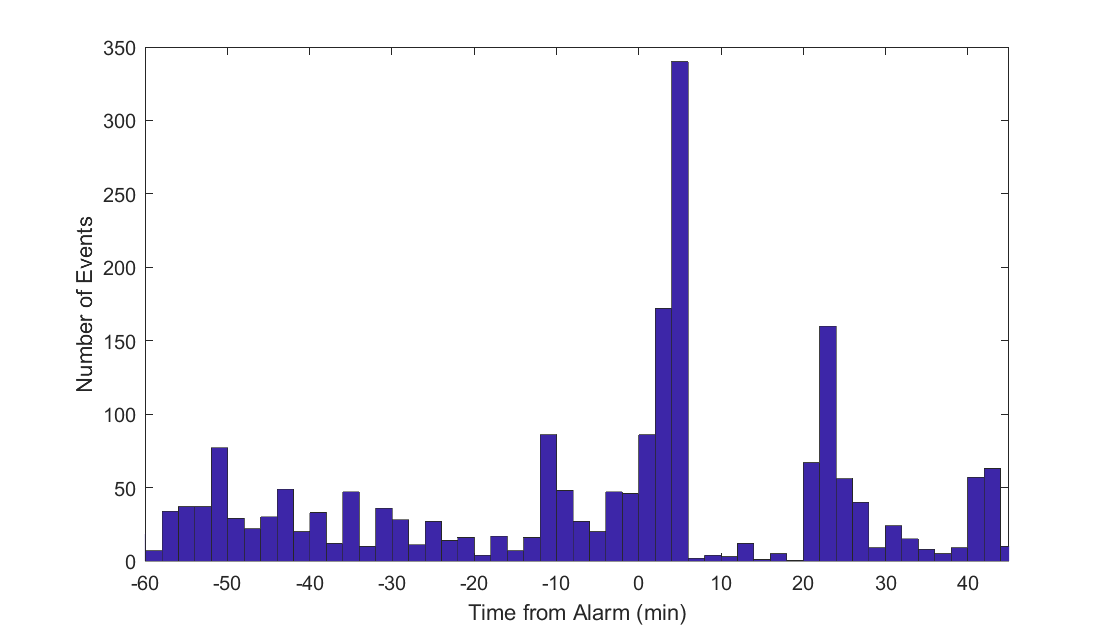}
\caption{\label{fig:Event_Histogram} Number of events detected and localized on the entire Fourth floor of Goodwin Hall for almost 2 hours surrounding the fire alarm. Detection and localization in this figure use vibration data which has been energy windowed using a large window, which is the proposed method for anonymized data.}
\end{figure}

In order to explore the usefulness of this level of localization accuracy to more real world data, all events happening on the fourth floor during the evacuation were also localized using a large window size. This process is similar to that which was performed to create Figure \ref{fig:ExitCounter}. When all events were detected and localized from the time before, during, and after the alarm, the amount of events over two minute windows is shown in Figure \ref{fig:Event_Histogram}. From Figure \ref{fig:Event_Histogram} it can easily be seen that just after the alarm sounds, the number of events goes up drastically. Information like this could be important for building management to understand how much utilization evacuation routes may experience during an emergency situation.

\begin{figure*}[!h]
\centering
\includegraphics[width = 6.2 in, height= 4 in]{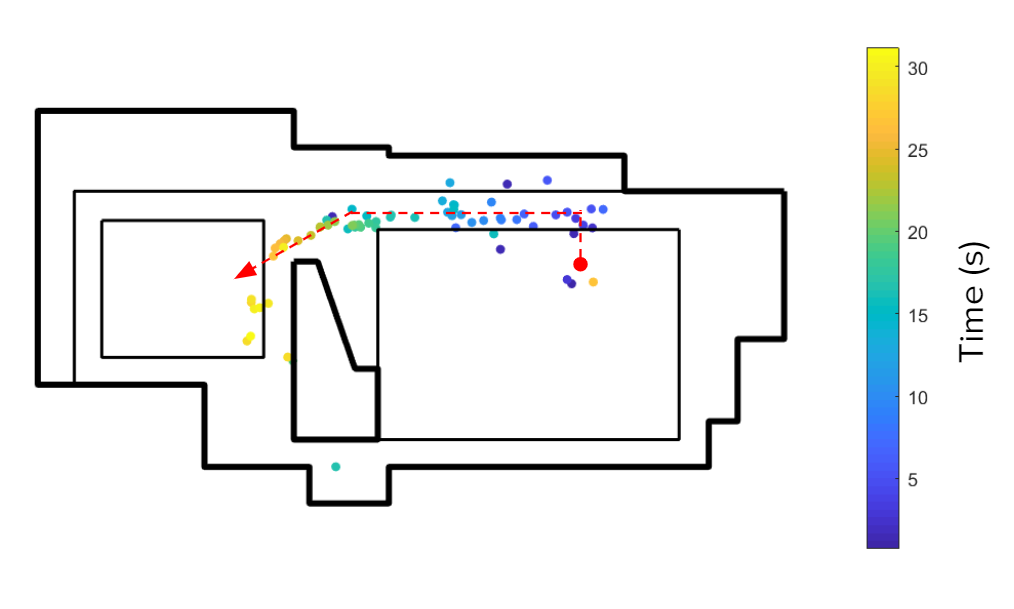}
\caption{\label{fig:EvacuationRoute} Localized footsteps from a single person evacuating the building. Each mark is a localized event, and marks are colored by their time within a roughly 30 second window. The overall path of the evacuation is shown in red.}
\end{figure*}

In order to get a more fine-grained view of how this localization is performing, a roughly 30 second subset of the fire-alarm dataset which corresponds to the single person walking signal shown in Figure \ref{fig:SingleEvacuation} was isolated. All events during this 30 seconds were localized, and plotted over a layout of the building, shown in Figure \ref{fig:EvacuationRoute}. Each event is also colored with the time that the event occurred in this 30 second time frame. There are a few outliers that do not follow the main pattern of the evacuation, which likely represent events from other people's actions during this time such as standing up from their desk or closing a door. Despite the outliers, and despite some spread from the localization error, the evacuation route taken by this individual can be clearly seen. The apparent evacuation route is highlighted with a red line in Figure \ref{fig:EvacuationRoute}. It appears that the individual got up from the desk in the office area by the red dot, exited into the hallway and walked down the hallway before turning a corner and entering a stairwell or getting on an elevator which are side by side near where the red arrow is shown.

Since the localization accuracy with large window sizes is around 1.4 meters, it is expected that we would see the spread in the localization that is visible in Figure \ref{fig:EvacuationRoute}. Despite this spread in the data, the evacuation route that the individual took is clearly visible. Once again, this shows that the proposed localization method is capable of generating results which are accurate enough to be useful on real-world data. Observing individual evacuation routes in a real, non-planned, alarm situation could be very useful data for building management professionals. With the data taken during this unplanned fire alarm, evacuation routes chosen by real individuals can be observed and compared to the nearest exits to those individuals, to ensure exit signs are clearly marked and evacuation procedures are being followed in the building. It could also be the case that due to the usage patterns of the building, one or more evacuation routes may be overused compared to others, and the evacuation planning for the building should be adjusted so that everyone in the building can be evacuated in an optimal manner.

\section{Discussion and Future Work}

\subsection{Classification and Gender Anonymization}

In this work, we trained binary classifiers that predicts the sex of an individual based on their footstep signature. Classifying the sex of the individual is a task representative of breaking the confidentiality of data. We use this task as a proof of concept, proving that we can preserve this confidentiality without losing the ability to locate an individual within a building. In particular, we use supervised machine learning to train an accurate binary classifier, aggregate the training data to an extent that sufficiently degrades the accuracy of the classifier, and we prove that this aggregated data is still sufficient for us to provide accurate localization. Even though we successfully prove our ability to preserve the anonymity of the participants that provided our data, our study had limitations due to the relatively small number of participants. This issue was overcome by generating a synthetic training data that has the same properties as the original training set. 

The challenges of the synthetic dataset set include its lack of diversity and its artificial bias reduction. The lack of diversity comes from the fact that it is generated from a dataset that has limited diversity. This lack of diversity is as a result of having only 16 participants. This means that even though we increase the number of participants, we do not learn individual characteristics that were missing from the original dataset. The reduced bias comes from the reduction in gender bias as a result of creating a dataset with an equal number of males and females. This approach was necessary because it gave us the ability to learn as much about female footstep signatures as we did about male ones. However, the reduction in gender bias might mean that our model may have missed some important parameters that are embedded within the gender bias inherent in real world data.

The challenges associated with the limited training data size lead to opportunities for future work. The reason, we had a small number of participants is that it is difficult to collect data for footstep signatures. Participants have to be recruited. Informed consent has to be provided. The building has to be empty, and this is challenging for large buildings.  Vibration readings have to be clean, pure or untainted, etc. An idea for future work is that of semi-supervised learning. Can one combine clean training data that has labels with unlabeled data that has some noise because it has been collected in conditions that are not ideal? For example, can one isolate a small portion of a building and collect footstep signature data while the rest of the building is occupied and use it for binary classification and localization by combining it with clean labelled data?

\subsection{Occupant Localization}

Similar to in our classification experiment, the localization had challenges associated with it. Our goal was to choose an energy based vibration localization method which would be compatible with our anonymization method, and investigate if localization was still possible with the anonymized data. Within the scope of this goal, we have shown that our localization method still functions well with our anonymization approach. Further work with this localization method could explore more avenues for optimizing our localization in an effort to get the errors more in line with other vibration based methods as shown in Table \ref{tab:VibrationBased}. One further parameter that can be optimized is the amount of time over which the energy of an event is counted, and how this interacts with the window size over which we are aggregating our data. For example with a small window size used for anonymization, multiple windows are currently added to get the total energy of the event, but how many is optimal? Another parameter which could be explored is the number of sensors used when fitting to obtain a location estimate. Currently our methodology has manually zoned the fourth floor into zones, which only localize within their own convex hulls. This is done to keep noise from far away from an event (possibly other events happening on the other side of the building) from interfering with the localization results. Another possibility would be to choose sensors only near the sensor which contains the maximum energy, possibly the closest $N$ sensors, or sensors within a certain distance. A final parameter is the possibility of trying to normalize our noise. Some sensors in the building have more inherent noise in them, from nearby machinery or HVAC systems. Including a method for separating energy from these building based noise sources from the energy of an actual footstep event could also increase localization energy.

One thing we encountered during our localization experiment is the challenge in localizing multiple events in close proximity and quick succession. We saw best results with a single person in an area as shown in Figure \ref{fig:EvacuationRoute}. However when multiple people were walking together while evacuating, it was more difficult to get reliable location estimates. There are two main challenges that could be explored here: separating sources and localizing multiple people at a time, or switching to track groups instead. If it is known that there are two people walking in the same area, it could be possible to use a small energy window to still get information enough to localize both people. Another method would be to use large windows, and instead of looking for individual events localize overall trends in events such as seeing a whole group of people move down a hallway. Each of these are worth further investigation.

Finally, future work for vibration based building localization in general is to demonstrate its usefulness in the real world. Multiple papers, including our work here, have shown different methods for locating people within a building. However, it is usually only speculated as to the usefulness of this data. We have endeavoured to show this to a degree, by tracking an individual's evacuation route during an unplanned alarm situation. This gives a real world example of how building localization could be used to inform building evacuation planning. Of further interest could be using real time localization data to control building climate control, in order to save energy. Another future work could also simulate tracking individuals in a simulated security emergency situation, so that police or first responders know exactly where in the building to focus their efforts.

\section{Conclusion}

In this work we have surveyed various methods used to track people within buildings, and shown that all current methods present privacy concerns for building users to be identified by the same data used to locate them. We then proposed a method for tracking people in a building while preserving their privacy, by using windowed floor vibration data. As the time window over which the vibration data is windowed increases, classifiers for an individuals gender are unable to classify better than 50 percent, showing that the individual's anonymity is preserved. Over these same time window sizes, the localization accuracy of the proposed method increases from an error of 1.1 m to 1.4 m. Combining these two results, we have shown that our proposed method can preserve the anonymity of individuals in a building while sacrificing little localization accuracy.




\appendices
\section{Author Contributions}

All authors were involved in all aspects of the project. Each author had a main focus, listed below.

\subsection{Ellis Kessler}

Ellis has been the main point of contact for the case study in Goodwin Hall. Ellis has been in charge of locating and cleaning previous data from Goodwin Hall to be adapted for this project. Ellis has also carried out the localization experiments and written the related work section on vibration based localization, and worked on the related work section.

\subsection{Moeti Masiane}

Moeti has worked on data pre-processing, visualization, and classifier training. Moe has conducted the exploratory data analysis and visualization. In addition to writing part of the related works section, Moe has also focused on training classification models for gender classification before and after anonymizing the data.

\subsection{Awad Abdelhalim}

Awad conducted the literature review on ethical concerns. Awad has reviewed and synthesized the literature on traditional tracking approaches, highlighting the development, ethical and privacy concerns associated with location-based and vision-based tracking technologies. He has assisted in analyzing the case-study data to ensure those concerns identified are addressed.



%



\bibliographystyle{ieeetr}
\bibliography{refs}

\end{document}